\documentclass[floatfix,preprint,authoryear]{elsarticle}
\usepackage{amsmath, amsthm, amssymb}
\usepackage{lineno,graphicx}

\journal{Journal of Theoretical Biology}

\begin{document}

\def\g{\gamma}
\def\r{\rho}
\def\w{\omega}
\def\wo{\w_0}
\def\wp{\w_+}
\def\wm{\w_-}
\def\t{\tau}
\def\av#1{\langle#1\rangle}
\def\pf{P_{\rm F}}
\def\pr{P_{\rm R}}
\def\F#1{{\cal F}\left[#1\right]}

\begin{frontmatter}

\title{A theoretical approach to understand spatial organization in complex ecologies}

\author[har,vtp]{Ahmed Roman}

\author[boo]{Debanjan Dasgupta}

\author[vtp,vtcsmbp,vtais]{Michel Pleimling\corref{cor1}}
\ead{pleim@vt.edu}

\cortext[cor1]{Corresponding author}

\address[har]{Department of Organismic and Evolutionary Biology, Harvard University, Cambridge, MA 02138, USA}
\address[vtp]{Department of Physics, Virginia Polytechnic Institute and State University, Blacksburg, Virginia 24061-0435, USA}
\address[boo]{Booz Allen Hamilton, 575 Herndon Parkway, Herndon, Virginia 20170, USA}
\address[vtcsmbp]{Center for Soft Matter and Biological Physics, Virginia Polytechnic Institute and State University, Blacksburg, Virginia 24061-0435, USA}
\address[vtais]{Academy of Integrated Science, Virginia Polytechnic Institute and State University, Blacksburg, Virginia 24061-0405, USA}

\date{\today}

\begin{abstract}
Predicting the fate of ecologies is a daunting, albeit extremely important, task. As part of this task one needs to develop an
understanding of the organization, hierarchies, and correlations among the species forming the ecology. Focusing on complex food networks
we present a theoretical method that allows to achieve this understanding. Starting from the adjacency matrix the method derives specific matrices 
that encode the various inter-species relationships. The full potential of the method is achieved in a spatial setting where
one obtains detailed predictions for the emerging space-time patterns. For a variety of cases these theoretical predictions are verified
through numerical simulations.
\end{abstract}


\end{frontmatter}


\section{Introduction}
Understanding the spatial structure of ecological networks is vital in theoretical and experimental biology and 
ecology \citep{May74,Smith74,Sole06,Smith82,Hof98,Now06,Weber14}. The stability of the dynamics of ecological networks 
is influenced by a variety of factors such as the topology of the network as well as the weights of the links composing the network 
(see \citep{Knebel13} for a recent example of Lotka-Volterra networks). A spatial setting, where individuals interact 
locally and are mobile, strongly influences the 
dynamics of the system and also has a large impact on the stability of an ecological network \citep{Rei08,Roman13}. 

The introduction of space gives rise to rich, interaction network dependent, phenomena including pattern formation, 
cluster coarsening, alliance formation and nested ecological niches.
These phenomena have yielded a huge theoretical interest recently, and
a variety of techniques \citep{Sza07,Fre09} have been employed to study these intriguing phenomena 
for symmetric networks, ranging from simple cases with three \citep{Rei08,Rei07,Pel08,Rei08a,Ven10,Shi10,Wan10,
He10,Win10,He11,Rul11,Wan11,Nah11,Jia11,He12,Juu12,Jia12,Lam12,Ada12,Juu13,Rul13,Szs13,Sch13,Szs14,Gro15}
and four species \citep{Rom12,Sza04,Sza07b,Sza08,Int13,Gui13,Int15}
to complex situations with an 
arbitrary number of species \citep{Roman13,Sza01,Sza01b,Sza05,Per07,Sza08a,Ave12a,Ave12b,Ave14,Mowlaei14,Sza07b,
Vuk13,Kan13,Ave14b,Che14}. Realistic ecologies, however, are endowed 
with complex interaction networks that can not be captured fully by only considering symmetric networks.
As such it is important to develop theoretical approaches that allow to understand the dynamics of 
general networks \citep{Sza07c,Lut12,Pro99,Van12,Knebel13,Dob14,Rul14,Var14,Sza15,Dal15,Szo15} and their effects on biodiversity, 
correlations, and spatio-temporal patterns.


In the following we present an exact method that allows to reveal why species are partitioned in domains and what are the partition sets of the species labels in complex food networks.
This information is a necessary first step in order to develop a predictive theory for the fate of ecologies. Our approach
predicts the alliance formation between different species and fixes the emerging space-time patterns in many spatial 
interaction networks. As a consequence, our approach also permits to shed light on the stability of ecological niches by breaking down this stability question into two distinctive parts: the stability of interactions among domains and within domains.    

\section{Model and examples}
\begin{figure} [!h]
 \begin{center}
 
~~~~~~(a)~~~~~~~~~~~~~~~~~~~~~~~~~~~~~~~~~~~(b)~~~~~~~~~\\
\includegraphics[width=0.35\columnwidth]{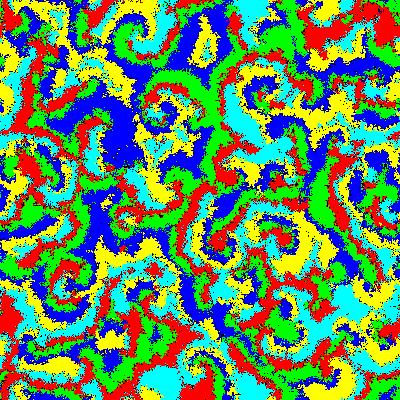}~~~
\includegraphics[width=0.35\columnwidth]{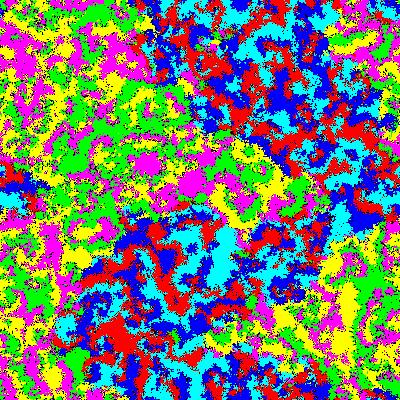}~~~\\[0.2cm]
~~~~~~(c)~~~~~~~~~~~~~~~~~~~~~~~~~~~~~~~~~~~(d)~~~~~~~~~\\
\includegraphics[width=0.35\columnwidth]{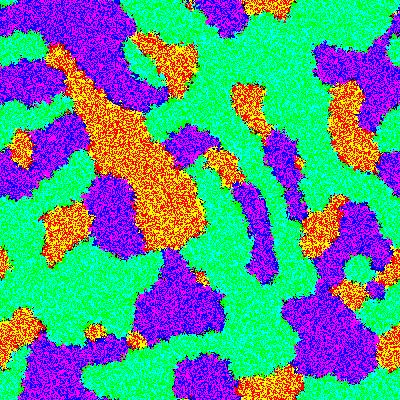}~~~
\includegraphics[width=0.35\columnwidth]{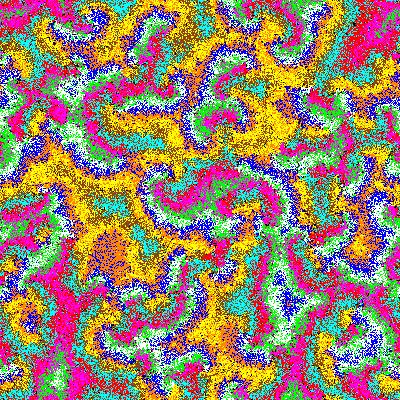}~~~
 \end{center}
\caption{(Color online) Four typical types of patterns observed in many species food networks (in all cases the rates where
chosen to be species independent): (a)
compact spirals in the $(5,3)$ model (with $\gamma = \delta =1$ and $\alpha = \beta =0$); (b) coarsening with spirals forming inside the domains in
the $(6,3)$ model (with $\gamma = \delta =1$ and $\alpha = \beta =0$); (c) coarsening of neutral species domains in the $(6,2)$ model
(with $\gamma = \delta =1$ and $\alpha = \beta =0$); (d) fuzzy spirals with mixing of species inside the spirals in the
$(9,3)$ model (with $\gamma = \delta = 0.5$ and $\alpha = \beta =0.25$).
Systems of $400 \times 400$ sites have been simulated.
\label{fig1}
}
\end{figure}

Fig.\ \ref{fig1} shows some typical two-dimensional space-time patterns that we aim at predicting with the approach discussed in the following.
Whereas snapshot (a) shows an example of a system with spirals where every wavefront contains only one species, panel (b) gives an
example of a coarsening process with two types of domains where inside every domain spirals are formed, thus yielding non-trivial
dynamics inside the coarsening domains. Example (c) reveals a coarsening process where every domain is formed by an alliance
of mutually neutral species. Finally, snapshot (d) shows a case of fuzzy spirals due to the mixing of different
species inside the waves. 

In order to focus our discussion we consider in the following systems of $N$ species in two space dimensions where 
individuals of different species have mutual predator-prey
interactions, give birth to off-springs, and are mobile. For our method
the details of the implementation of these interactions do not matter, with the exception that the predation-prey interactions should
be binary, i.e. at any given event one predator consumes one prey. For the examples in Fig. \ref{fig1} as well as for the more complicated cases
discussed later we consider the following general interaction scheme ($X_i$ denotes a member of species $i$ and
${\emptyset}$ is an empty site):
\begin{eqnarray} 
X_{i} + X_{j}& \xrightarrow{\delta_{ij}} & {\emptyset}+X_{i}\label{model1} \\
X_{i} + X_{j}& \xrightarrow{\alpha_{ij}} & X_{j}+X_{i} \label{model2} \\
X_{i} + {\emptyset}& \xrightarrow{\gamma_{i}} & X_{i}+X_{i} \label{model3} \\
X_{i} + {\emptyset}& \xrightarrow{\beta_{i}} & {\emptyset}+X_{i} \label{model4} 
\end{eqnarray}
where every lattice site can be occupied by at most one individual, whereas reactions between different individuals only
occur when they are at neighboring sites. Our approach remains valid when considering in addition self-predation.
The model parameters $\delta_{ij}$, $\alpha_{ij}$, $\gamma_{i}$, and $\beta_{i}$
are the rates at which these reactions take place. 
For the examples included in this paper, predation and swapping rates are always considered
to be independent of the prey, i.e. $\delta_{ij} = \delta_i$ and $\alpha_{ij} = \alpha_i$.
The separation of the predation and birth events is needed for the formation of spirals as those shown in Fig. \ref{fig1},
but is otherwise not relevant in order to establish the relationships between species.
We also allow for various ways of mobility (swapping of individuals
on neighboring sites (\ref{model3}) or diffusion by jumping to empty sites (\ref{model4})), in order to be as general as possible.

The reaction scheme (\ref{model1})-(\ref{model2}) contains many known situations as for example the cyclic $N$ species Lotka-Volterra and
May-Leonard models as well as more complicated situations like the $(N,r)$ models where each species preys on $r$ other species 
in a cyclic way \citep{Roman13,Mowlaei14}. Whereas the examples shown in Fig. \ref{fig1} are those of $(N,r)$ models with 
symmetric rates (hence we can drop the labels $i$, $j$ for the rates, see the figure caption), more general cases
will be discussed below. 

The matrix approach discussed in the following is valid for rather general predator-prey networks, provided that
no interactions of higher order than second order are present. We assume that the species network gives rise
to a unique partition of the species into domains. While we have not encountered a case where this assumption is
not fulfilled, we can not exclude that there exist networks of species that exhibit in presence of noise two non-isomorphic partitions 
of species into domains. For the cases used to illustrate our approach, we consider rates for which none 
of the species go extinct before the quasi-steady state of the dynamics is reached, i.e.
the time scale on which the species are forming alliances and domains is shorter than the extinction time
for any of the species. Notably this requires that the diffusion and mobility rates are not too high so as to induce rapid extinction
of the species, but are still large enough to allow the formation of species alliances.

\section{Matrix approach}
The predation scheme (\ref{model1}) can be represented by a graph where one connects species $i$ and $j$ by a directed edge 
if $i$ preys on $j$. The scheme can also be encoded in the $N \times N$ {\it adjacency matrix} $\bold{\underline{A}}$ where the element
$a_{ij} = 1$ if $i$ preys on $j$ and 0 otherwise. The {\it square of the adjacency matrix} $\bold{\underline{B}} = \bold{\underline{A}}^2$
contains information about preferred partnership formations based on the maximum protection a species can enjoy while being
close to some other species. Indeed, the element $b_{ij}$ counts the number of
directed paths of length 2 from vertex $i$ to vertex $j$, i.e. the number of paths of the form $i \longrightarrow k \longrightarrow j$
where $k$ is a vertex different from $i$ and $j$. Following the maxim that "the enemy of my enemy is my friend," species $j$
wants to be in close proximity to that species that attacks most of its predators. Mathematically, this preferred partner of
$j$ is obtained from the condition $\max_i b_{ij}$.  

Of course, a given species does not want to undergo a close partnership with one of its predators. Adding this important information
yields the {\it alliance matrix} $\bold{\underline{S}}$ whose elements $s_{ij}$ provide information on the best possible allies $j$
for each species $i$:
\begin{equation}
s_{ij} = \delta_{b_{ji}, m_i}\delta_{0,a_{ji}}~,
\end{equation}
where $m_i =  \max_{j} b_{ji}$. The first condition selects those species $j$ that are preferred partners of $i$ as they
provide the best protection by attacking the largest number of its predators, whereas the second condition makes sure that $j$ is not itself a predator of $i$, yielding 
a matrix with elements $s_{ij} = 1$ for the best possible allies of species $i$ and zero otherwise.

In order to fully understand space-time patterns as those shown in Fig. \ref{fig1}, one needs to further distinguish between preferred allies
of species $i$ that are hunted by species $i$ and preferred allies towards which species $i$ has a neutral approach. This information is encoded by the two matrices
$\bold{\underline{P}}$ ({\it prey-allies}) and $\bold{\underline{N}}$ ({\it neutral allies}) with elements
\begin{eqnarray}
p_{ij} &=& s_{ij} \wedge a_{ij}=\delta_{s_{ij},a_{ij}} \\
n_{ij} &=& s_{ij }\wedge \neg \, a_{ij}=\delta_{s_{ij},(1-a_{ij})}
\end{eqnarray}
with the binary operators AND ($\wedge$) and NOT ($\neg$).

The two matrices $\bold{\underline{P}}$ and $\bold{\underline{N}}$ allow to understand (and predict) many of the space-time patterns that can emerge
in predator-prey systems governed by the reactions (\ref{model1})-(\ref{model2}). A matrix element $p_{ij}=1$ means that species $i$ seeks
the protection of species $j$, while at the same time preying on species $j$. As a result single species waves will form where a wave containing
individuals of species $i$ will follow one formed by species $j$. On the other hand $n_{ij}=1$ represents non-aggressive coexistence
where species $i$ is protected by a neutral species $j$ (i.e. $i$ does not prey on $j$). As a result $i$ will mix with $j$ in order
to enjoy the enhanced protection by its preferred ally. If $n_{ij}=n_{ji}=1$, then the two mutually neutral partners will form neutral domains
and protect each other.

Inspection of Fig. \ref{fig1}d reveals a case that is not yet covered by the matrices given above, namely that of 'fuzzy' waves where
three species mix to some degree. Consider three given species $i$, $j$, and $k$. The above
 situation is encountered when species $i$ and $j$ are mutually neutral and species $j$ and $k$ are mutually neutral,
while at the same time species $i$ preys on species $k$. This means that the following conditions have to be fulfilled simultaneously:
$n_{ij}=1$, $n_{jk}=1$, and $a_{ik}=1$. As a result, $i$ chases species $k$ while both mix with the intermediate species $j$, giving fuzzy
looking waves as those seen in Fig. \ref{fig1}d. Defining the matrix elements of a matrix $\bold{\underline{F}}$ as
\begin{equation} \label{eq:f}
f_{ij}=\theta \left(\left(  \sum_{k=1}^{N} \delta_{n_{ij},1} \delta_{n_{jk},1} \delta_{a_{ik},1} \right) - 1\right)~,
\end{equation}
with $\theta(x) = 1$ for $x \ge 0$ and zero otherwise,
we have that $f_{ij}=1$ if $j$ acts as a neutral intermediary between species $i$ and some other species, whereas $f_{ij}=0$ otherwise. 

\section{Examples}
Let us first look at the four cyclic $(N,r)$ games of Fig. \ref{fig1} in order to see how to work with the different matrices.

For the $(5,3)$ game shown in Fig. \ref{fig1}a the relevant matrix is
\begin{equation}
\renewcommand{\arraystretch}{.45}
\bold{\underline{P}}=\left(
\begin{array}{ccccc}
 0 & 1 & 0 & 0 & 0 \\
 0 & 0 & 1 & 0 & 0 \\
 0 & 0 & 0 & 1 & 0 \\
 0 & 0 & 0 & 0 & 1 \\
 1 & 0 & 0 & 0 & 0 \\
\end{array}
\right)
\end{equation}
whereas $\bold{\underline{N}} = \bold{\underline{F}} = \bold{\underline{0}}$. We readily obtain from the prey-allies matrix
$\bold{\underline{P}}$ that species 1 wants to ally with its prey species 2, species 2 wants to ally  
with its species 3, etc. This yields single species spirals where the species follow each other in the order
$1 \rightarrow 2 \rightarrow 3 \rightarrow 4 \rightarrow 5 \rightarrow 1$.

The $(6,3)$ in Fig. \ref{fig1}b is characterized by a similar looking $6 \times 6$ matrix
\begin{equation}
\renewcommand{\arraystretch}{.45}
\bold{\underline{P}}=\left(
\begin{array}{cccccc}
0 & 0 & 1 & 0 & 0 & 0 \\
 0 & 0 & 0 & 1 & 0 & 0 \\
 0 & 0 & 0 & 0 & 1 & 0 \\
 0 & 0 & 0 & 0 & 0 & 1 \\
 1 & 0 & 0 & 0 & 0 & 0 \\
 0 & 1 & 0 & 0 & 0 & 0 \\
\end{array}
\right)
\end{equation}
where again $\bold{\underline{N}} = \bold{\underline{F}} = \bold{\underline{0}}$. This time, however, two family of spirals
(or teams) are formed, one with the ordering $1 \rightarrow 3 \rightarrow 5 \rightarrow 1$, the other with the ordering 
$2 \rightarrow 4 \rightarrow 6 \rightarrow 2$. As every member in a team has both a prey and a predator in the other team,
these two family of spirals compete against each other, which yields a coarsening process with a three-species rock-paper-scissors
game within each domain.

The $(6,2)$ case of Fig. \ref{fig1}c is a case where the prey-allies matrix $\bold{\underline{P}}$ vanishes whereas the neutral allies matrix
$\bold{\underline{N}}$ is non-zero:
\begin{equation}
\renewcommand{\arraystretch}{.45}
\bold{\underline{N}}=\left(
\begin{array}{cccccc}
 0 & 0 & 0 & 1 & 0 & 0 \\
 0 & 0 & 0 & 0 & 1 & 0 \\
 0 & 0 & 0 & 0 & 0 & 1 \\
 1 & 0 & 0 & 0 & 0 & 0 \\
 0 & 1 & 0 & 0 & 0 & 0 \\
 0 & 0 & 1 & 0 & 0 & 0 \\
\end{array}
\right)~.
\end{equation}
Inspection of $\bold{\underline{N}}$ reveals the formation of three teams composed of two mutually neutral species (1 and 4, 
2 and 5, 3 and 6), resulting in a coarsening process with three different types of domains.

Finally, the $(9,3)$ game in Fig. \ref{fig1}d provides us with an example where both $\bold{\underline{N}}$ and $\bold{\underline{F}}$
are non-zero:
\begin{equation}
\renewcommand{\arraystretch}{.45}
\bold{\underline{N}}=\bold{\underline{F}}=
\left(
\begin{array}{ccccccccc}
 0 & 0 & 0 & 0 & 0 & 1 & 0 & 0 & 0 \\
 0 & 0 & 0 & 0 & 0 & 0 & 1 & 0 & 0 \\
 0 & 0 & 0 & 0 & 0 & 0 & 0 & 1 & 0 \\
 0 & 0 & 0 & 0 & 0 & 0 & 0 & 0 & 1 \\
 1 & 0 & 0 & 0 & 0 & 0 & 0 & 0 & 0 \\
 0 & 1 & 0 & 0 & 0 & 0 & 0 & 0 & 0 \\
 0 & 0 & 1 & 0 & 0 & 0 & 0 & 0 & 0 \\
 0 & 0 & 0 & 1 & 0 & 0 & 0 & 0 & 0 \\
 0 & 0 & 0 & 0 & 1 & 0 & 0 & 0 & 0 \\
\end{array}
\right)~,
\end{equation}
whereas $\bold{\underline{P}} = \bold{\underline{0}}$.
It follows from this that species 5, which is the preferred ally of species 1, acts as an intermediary between species 1 and 9. As species 9 preys
on species 1, this yields a wavefront of the type $9 \stackrel{\left[ 5 \right]}{\longrightarrow} 1$ where $\left[ \cdots \right]$ indicates
a neutral intermediary. As this happens in a cyclic way, one observes the emergence of fuzzy spirals, with the ordering
$1 \stackrel{\left[ 6 \right]}{\longrightarrow} 2 \stackrel{\left[ 7 \right]}{\longrightarrow} 3 
\stackrel{\left[ 8 \right]}{\longrightarrow} 4 \stackrel{\left[ 9 \right]}{\longrightarrow} 5 \stackrel{\left[ 1 \right]}{\longrightarrow} 6 
\stackrel{\left[ 2 \right]}{\longrightarrow} 7 \stackrel{\left[ 3 \right]}{\longrightarrow} 8 \stackrel{\left[ 4 \right]}{\longrightarrow}
9 \stackrel{\left[ 5 \right]}{\longrightarrow} 1$.

\begin{figure} [!h]
 \begin{center}

~~~~~~~~~$t=100$~~~~~~~~~~~~~~~~~~~~~~~~~~~~~~~$t=900$~~~~~~~~~~\\
\includegraphics[width=0.35\columnwidth]{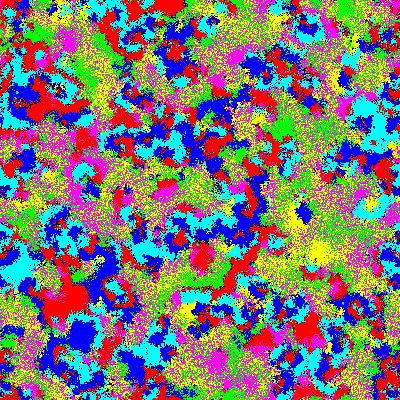}~~~
\includegraphics[width=0.35\columnwidth]{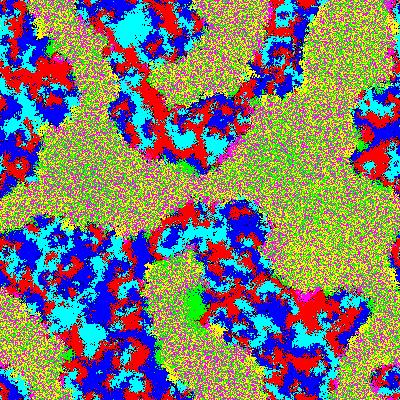}~~~
%

 \end{center}
\caption{(Color online)
Six-species model defined by the adjacency matrix (\ref{A_example1}) and the rates (\ref{R_example1}).
Red: species 1, green: species 2, blue: species 3, yellow: species 4, cyan: species 5, pink: species 6.
The system size is $500 \times 500$.
Times are measured since preparing the system, where one time step corresponds to $500 \times 500$ 
proposed updates.
\label{fig2}
}
\end{figure}

\begin{figure} [!h]
 \begin{center}

~~~~~~~~~$t=200$~~~~~~~~~~~~~~~~~~~~~~~~~~~~~~~$t=600$~~~~~~~~~~\\
\includegraphics[width=0.35\columnwidth]{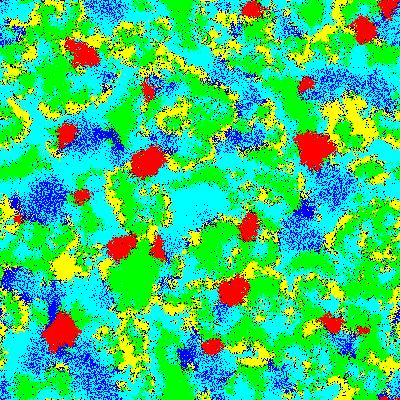}~~~
\includegraphics[width=0.35\columnwidth]{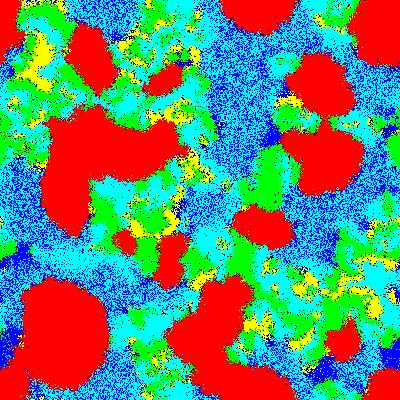}~~~
%

 \end{center}
\caption{(Color online)
Five-species model defined by the adjacency matrix (\ref{A_example2}) and the rates (\ref{R_example2}).
Red: species 1, green: species 2, blue: species 3, yellow: species 4, cyan: species 5.
The system size is $400 \times 400$.
Times are measured since preparing the system, where one time step corresponds to $400 \times 400$
proposed updates.
\label{fig3}
}
\end{figure}

Let us now look at the two more involved examples shown in Figs. \ref{fig2} and \ref{fig3} 
where the interaction scheme is no longer cyclic.

The snapshots in Fig.\ \ref{fig2} result from a six-species model with an interaction scheme described by the matrix
\begin{equation} \label{A_example1}
\renewcommand{\arraystretch}{.45}
{\bold{\underline{A}}}= \left(
\begin{array}{cccccc}
 0 & 1 & 1 & 1 & 0 & 0 \\
 0 & 0 & 1 & 0 & 1 & 0 \\
 0 & 0 & 0 & 1 & 1 & 1 \\
 1 & 0 & 0 & 0 & 1 & 0 \\
 1 & 1 & 0 & 0 & 0 & 1 \\
 1 & 0 & 1 & 0 & 0 & 0 \\
\end{array}
\right)
\end{equation}
and rates given by
\begin{equation} \label{R_example1}
\renewcommand{\arraystretch}{.45}
\left(
\begin{array}{cccc}
\delta_1 & \alpha_1 & \gamma_1 & \beta_1 \\
\delta_2 & \alpha_2 & \gamma_2 & \beta_2 \\
\delta_3 & \alpha_3 & \gamma_3 & \beta_3 \\
\delta_4 & \alpha_4 & \gamma_4 & \beta_4 \\
\delta_5 & \alpha_5 & \gamma_5 & \beta_5 \\
\delta_6 & \alpha_6 & \gamma_6 & \beta_6 
\end{array}\right)=
\left(
\begin{array}{cccc}
0.9 & 0.08 & 0.8 & 0.16 \\
0.43 & 0.46 & 0.8 & 0.16 \\
0.9 & 0.08 & 0.8 & 0.16 \\
0.43 & 0.46 & 0.8 & 0.16 \\
0.9 & 0.08 & 0.8 & 0.16 \\
0.43 & 0.46 & 0.8 & 0.16 
\end{array}\right).
\end{equation}
Following the construction explained above, we get the matrices
\begin{equation}
\renewcommand{\arraystretch}{.45}
{\bold{\underline{P}}}= \left(
\begin{array}{cccccc}
 0 & 0 & 1 & 0 & 0 & 0 \\
 0 & 0 & 0 & 0 & 0 & 0 \\
 0 & 0 & 0 & 0 & 1 & 0 \\
 0 & 0 & 0 & 0 & 0 & 0 \\
 1 & 0 & 0 & 0 & 0 & 0 \\
 0 & 0 & 0 & 0 & 0 & 0 \\
\end{array}
\right)
~;~
{\bold{\underline{N}}}= \left(
\begin{array}{cccccc}
 0 & 0 & 0 & 0 & 0 & 0 \\
 0 & 0 & 0 & 1 & 0 & 0 \\
 0 & 0 & 0 & 0 & 0 & 0 \\
 0 & 0 & 0 & 0 & 0 & 1 \\
 0 & 0 & 0 & 0 & 0 & 0 \\
 0 & 1 & 0 & 0 & 0 & 0 \\
\end{array}
\right)
\end{equation}
which, together with the fact that ${\bold{\underline{F}}} = {\bold{\underline{0}}}$,
allow us to predict the alliances and space-time pattern appearing in this system. In the present case,
both the prey-allies and neutral allies matrices are non-zero, so that we can expect the formation of
both neutral alliances as well as of alliances where the partners remain in a predator-prey relationship.
Indeed, from the matrix ${\bold{\underline{P}}}$ we obtain that species 1, 3, and 5 form a three-species
cyclic game that results in the formation of compact wavefronts with the following ordering of
the species: $1 \rightarrow 3 \rightarrow 5 \rightarrow 1$. On the other hand, from the matrix ${\bold{\underline{N}}}$
follows that the species 2, 4, and 6 form a neutral alliance. As a result, two types of domains should form,
one composed of three neutral partners and one composed of three partners that undergo a cyclic rock-paper-scissors
game, resulting in a coarsening process with different internal dynamics inside the domains.

The five-species model shown in Fig. \ref{fig3} is defined by the adjacency matrix
\begin{equation} \label{A_example2}
\renewcommand{\arraystretch}{.45}
{\bold{\underline{A}}}= \left(
\begin{array}{ccccc}
 0 & 1 & 1 & 1 & 1 \\
 1 & 0 & 1 & 1 & 0 \\
 1 & 0 & 0 & 1 & 0 \\
 1 & 0 & 0 & 0 & 1 \\
 1 & 1 & 0 & 0 & 0
\end{array}\right)~.
\end{equation}
For the time evolution shown in Fig. \ref{fig3} we used the rates
\begin{equation} \label{R_example2}
\renewcommand{\arraystretch}{.45}
\left(
\begin{array}{ccccc}
\delta_1 & \alpha_1 & \gamma_1 & \beta_1 \\
\delta_2 & \alpha_2 & \gamma_2 & \beta_2 \\
\delta_3 & \alpha_3 & \gamma_3 & \beta_3 \\
\delta_4 & \alpha_4 & \gamma_4 & \beta_4 \\ 
\delta_5 & \alpha_5 & \gamma_5 & \beta_5 
\end{array}\right)=
\left(
\begin{array}{ccccc}
0.5 & 0.4 & 0.8 & 0.16 \\
0.7 & 0.24 & 0.8 & 0.16 \\
0.3 & 0.56 & 0.8 & 0.16 \\
0.8 & 0.16 & 0.8 & 0.16 \\
0.3 & 0.56 & 0.8 & 0.16 
\end{array}\right).
\end{equation}
From ${\bold{\underline{A}}}$ we readily obtain
\begin{equation}
\renewcommand{\arraystretch}{.45}
{\bold{\underline{P}}}= \left(
\begin{array}{ccccc}
 0 & 0 & 0 & 0 & 0 \\
 0 & 0 & 0 & 1 & 0 \\
 0 & 0 & 0 & 0 & 0 \\
 0 & 0 & 0 & 0 & 1 \\
 0 & 1 & 0 & 0 & 0
\end{array}\right)
~;~
{\bold{\underline{N}}}= \left(
\begin{array}{ccccc}
 1 & 0 & 0 & 0 & 0 \\
 0 & 0 & 0 & 0 & 0 \\
 0 & 0 & 0 & 0 & 1 \\
 0 & 0 & 0 & 0 & 0 \\
 0 & 0 & 1 & 0 & 0
\end{array}\right)~,
\end{equation}
whereas ${\bold{\underline{F}}}={\bold{\underline{0}}}$ (i.e. no fuzzy wavefronts are formed). The matrices
immediately yield all the relevant information regarding the partnerships forming in this system. As
$n_{11} =1$, species 1 wants to self-segregate, which is a direct consequence of the fact that all the
other species are both its predators and preys. The neutral allies matrix ${\bold{\underline{N}}}$ also indicates that species 3 and 5
form a neutral alliance, in order to fight off their corresponding predators. Finally, the prey-allies matrix ${\bold{\underline{P}}}$
reveals that the three species 2, 4, and 5 tend to form an active three-species alliance with the following ordering of the wave fronts:
$2 \rightarrow 4 \rightarrow 5 \rightarrow 2$. Our method therefore predicts the formation of three types of domains (pure species 1 domains, domains
of the neutral alliance between species 3 and 5, as well as domains where species 2, 4, and 5 undergo a cyclic game), where, remarkably,
species 5 is involved in {\it two} different alliances. All these predictions are indeed verified when inspecting the snapshots in
Fig. \ref{fig3}.

Knowing the partnerships and the resulting space-time pattern is an important step in order to better understand
the fate of an ecology. This alone, though, is not enough to predict whether a given species will go extinct, as the
survival of a species also depends on the rates of its interactions with the other species. The reliability of our predictions at least for the $(N,r)$ model with symmetric rates is confirmed via exact agreement with the theoretical predictions in \citep{Mowlaei14} obtained using the complex Ginzburg Landau approach. Comprehensive numerical studies are required to fully understand the parameter regimes for which the presented ad-hoc methods are valid. We leave such calculations for future investigations.    

\section{Summary}
We have presented a theoretical approach that allows to predict partnership formation and emerging space-time patterns in
many-species food networks. While our emphasis was on predator-prey systems, the proposed use of various matrices, derived from
the adjacency matrix, to understand alliances and spatio-temporal pattern formation can be straightforwardly expanded to
other systems, ranging from ecology to social sciences \citep{Smith82,Hof98,Now06,Hau02,Gok14}, 
governed by binary interactions between different 'species'.


\section*{Acknowledgements}
M.P. thanks the Galileo Galilei
Institute for Theoretical Physics for the hospitality and
the INFN for partial support during the completion of
this work.\\
{\it Funding:} This work is supported by the US National
Science Foundation through grant DMR-1205309.

\end{document}